\documentclass[aps,prb,twocolumn,preprintnumbers,amsmath,amssymb]{revtex4-2}
\usepackage{lineno,hyperref,amsmath}
\usepackage{booktabs}
\usepackage{graphicx}% Include figure files
\usepackage{subfigure}
\usepackage{float}
\usepackage{amsmath}
\usepackage{amssymb}
\usepackage{appendix}
\usepackage{tikz}

\hypersetup{hidelinks,
colorlinks=true,
allcolors=blue,
pdfstartview=Fit,
breaklinks=true}

\definecolor{lime}{HTML}{A6CE39}
\DeclareRobustCommand{\orcidicon}{
\begin{tikzpicture}
\draw[lime, fill=lime] (0,0)
circle[radius=0.16]
node[white]{{\fontfamily{qag}\selectfont \tiny \.{I}D}};
\end{tikzpicture}
\hspace{-2mm}
}
\foreach \x in {A, ..., Z}{%
\expandafter\xdef\csname orcid\x\endcsname{\noexpand\href{https://orcid.org/\csname orcidauthor\x\endcsname}{\noexpand\orcidicon}}
}

\begin{document}

\title{Quasiparticle interference in altermagnets}

\author{Hao-Ran Hu}
\affiliation{National Laboratory of Solid State Microstructures, School of Physics, 
Jiangsu Physical Science Research Center, and Collaborative Innovation Center of Advanced Microstructures,
Nanjing University, Nanjing 210093, China}
\author{Xiangang Wan}
\affiliation{National Laboratory of Solid State Microstructures, School of Physics, 
Jiangsu Physical Science Research Center, and Collaborative Innovation Center of Advanced Microstructures,
Nanjing University, Nanjing 210093, China}
\author{Wei Chen \hspace{-1.5mm}\orcidA{}}
\email{Corresponding author: pchenweis@gmail.com}
\affiliation{National Laboratory of Solid State Microstructures, School of Physics, 
Jiangsu Physical Science Research Center, and Collaborative Innovation Center of Advanced Microstructures,
Nanjing University, Nanjing 210093, China}

\date{\today}

\begin{abstract}

A novel collinear magnetic phase, termed ``altermagnetism,'' has recently been {uncovered},
characterized by zero net magnetization and momentum-dependent collinear 
spin-splitting. To understand the intriguing physical effects of altermagnets and explore
their potential applications, it is crucial to analyze both the
geometric and spin configurations of altermagnetic Fermi
surfaces. Here, we conduct a comprehensive study of the quasiparticle interference (QPI) effects induced by both nonmagnetic and magnetic impurities 
in metallic altermagnets, incorporating the influence of Zeeman splitting and spin-orbit coupling. 
By examining the QPI patterns for various spin polarizations of magnetic impurities and different spin-probe channels, we identify a series of distinctive signatures that can be used to characterize altermagnetic Fermi surfaces. These predicted signatures can be directly compared with experimental results obtained through spin-resolved scanning tunneling spectroscopy.
\end{abstract}

\maketitle

\section{\label{sec1}Introduction}

Magnetism is an important and expansive research area of 
condensed matter physics, which is of significance for both fundamental science and device applications. 
Historically, two primary collinear magnetic phases are ferromagnetism 
and antiferromagnetism~\cite{Neel1971}. Very recently, a novel collinear magnetic phase 
called altermagnetism has been proposed, which has attracted increasing 
interest~\cite{Smejkal2022,Mazin2022,Smejkal2022b,Hayami2019}. What distinguishes altermagnet
from both ferromagnet and conventional antiferromagnet is the notable collinear momentum-dependent 
spin splitting despite zero net magnetization, which are induced by nonrelativistic spin and crystal rotational symmetry~\cite{Smejkal2022,Mazin2022,Smejkal2022a}. 
{Specifically, ferromagnets exhibit $\mathcal{T}$ (time-reversal)-symmetry breaking with momentum-independent spin splitting of bands, leading to isotropic $s$-wave spin-split Fermi surfaces. Antiferromagnets, on the other hand, maintain $\mathcal{T}$-invariant spin-degenerate bands that resemble those of nonmagnetic systems. In contrast, altermagnets feature $\mathcal{T}-$symmetry breaking spin splitting with alternating signs, along with anisotropic sublattice spin densities. These systems also exhibit more complex Fermi surface topologies, including anisotropic $d$-wave, $g$-wave, or $i$-wave spin-split Fermi surfaces~\cite{Smejkal2022,Mazin2022,Smejkal2022b,Hayami2019}.}
It leads to a range of fascinating physical effects, such as the crystal 
magneto-optical Kerr effect~\cite{Feng2015}, giant magnetoresistance~\cite{Smejkal2022a}, 
generation of spin currents~\cite{Shao2021,GonzalezHernandez2021,Bose2022}, 
and the anomalous Hall effect~\cite{Nagaosa2010,Sato2024,Hayami2021}, 
which are absent in traditional antiferromagnets. 
Due to its interesting properties and fascinating interplay with other physical scenarios, 
altermagnetism may stimulate research efforts in various branches of condensed matter physics,
such as spintronics \cite{Bai2023,Bai2022,Karube2022}, correlated matter states \cite{Smejkal2022}, 
and superconductivity \cite{Ouassou2023,Sun2023}. Hundreds of 
candidates for altermagnets have been predicted~\cite{Savitsky2024,Hayami2020}, including three-dimensional compounds like 
$\mathrm{MnTe}$, $\mathrm{RuO_2}$, and $\mathrm{La_2CuO_4}$ \cite{Smejkal2022,Mazin2023,FariaJunior2023,Smejkal2023}, 
as well as two-dimensional monolayers such as $\mathrm{RuF_4}$ with $d$-wave altermagnetic symmetry~\cite{Milivojevic2024} {and $\mathrm{V_2Se_2O}$ with $C$-pair spin-valley locking~\cite{Ma2021}}.
The database of altermagnetic materials is still rapidly expanding.

To fully understand the intriguing physical effects of altermagnets 
and explore their potential applications, it is essential to analyze 
the geometric and spin configurations of their Fermi surfaces. Recent 
studies utilizing spin-angle-resolved photoemission spectroscopy
have confirmed the altermagnetic lifting of Kramers spin degeneracy in 
$\mathrm{RuO_2}$ and $\mathrm{MnTe}$~\cite{Osumi2024,lin2024observationgiantspinsplitting}.
However, experiments on $\mathrm{RuO_2}$ using different approaches reveal 
a noticeable contradiction, as the muon spin rotation and relaxation experiment 
indicates the absence of magnetic order~\cite{Hiraishi2024,Kessler2024}. 
In this context, it is important to characterize altermagnetic features using 
additional methods such as quantum oscillation~\cite{Li2024}.
Another standard and powerful method for the characterization of Fermi surfaces as well
as their spin textures are the quasiparticle interference (QPI) induced by impurities,
which can be extracted by scanning tunneling spectroscopy (STS). It has been applied to
study the properties of Fermi surfaces in various quantum materials~\cite{Avraham2018}, 
including metals~\cite{Crommie1993,Sprunger1997}, 
superconductors~\cite{Capriotti2003,Wang2003,Hanaguri2009,Hoffman2002,PeregBarnea2003,Lee2009}, 
graphene~\cite{Rutter2007,PeregBarnea2008,Brihuega2008,Dombrowski2017,Jolie2018}, 
surface states of topological insulators~\cite{Zhou2009,Lee2009a,Beidenkopf2011,Kohsaka2015,Kohsaka2017}, 
Weyl semimetals~\cite{Batabyal2016,Inoue2016,Mitchell2016,Zheng2018}, and nonsymmorphic 
materials~\cite{Topp2017,Queiroz2018,Zhu2018}. 
{There are also studies on QPI patterns in other unconventional magnetic systems or materials with complex spin textures \cite{Biderang2022,Biderang2018,Hirschfeld2015,Akbari2013}.}
QPI is manifested by
the modulation of local density of states (LDOS) in space, which is then 
analyzed in the momentum space to provide useful information of the Fermi surfaces. 
When Fermi surfaces possess nontrivial spin textures, such as the 
present case of altermagnets, analyzing the spin-resolved modulation of LDOS
and its Fourier transform becomes essential, which can be achieved 
by spin-resolved STS. However, up until now,
the related theoretical and experimental investigations are still lacking.

In this work, we investigate the QPI in metallic altermagnets 
induced by nonmagnetic and magnetic impurities in different scattering and probe channels. 
The effects due to Zeeman splitting and spin-orbit coupling (SOC) 
are taken into account, which can be implemented in real materials. The standard 
T-matrix formalism is employed to calculate the QPI patterns. 
We further decompose the Fourier-transformed local density of states (FT-LDOS) into different types of response functions 
and spin-coherent factors using the Lehmann representation to analyze the resulting patterns.
		It is shown that the QPI patterns can faithfully reveal the 
geometric information and spin textures of the Fermi surface in altermagnets.
								Specifically, for the pristine $d$-wave altermagnet, QPI analysis directly reveals the altermagnetic Fermi surface, characterized by two ellipses with orthogonal major axes and oppositely oriented spins.
In the presense of 
a Zeeman field or the SOC effect, the QPI patterns dominated by the backscattering 
exhibit an petal-shaped contour, manifesting a Lifshitz transition of the 
Fermi surface. Moreover, analyzing the enhancement and suppression 
in various probe and scattering channels enables us to extract detailed 
spin information at the Fermi surfaces. 
We find that all the QPI patterns for the altermagnet subject to a Zeeman field are real, while the SOC case introduces imaginary components.
Therefore, by examining the imaginary parts of the FT-LDOS in certain channels, we can easily distinguish 
between the scenarios with Zeeman splitting and SOC. Through 
a combination of analytical and numerical calculations, we demonstrate that the modulations of the QPI 
patterns with varying physical conditions can effectively capture both the geometric and spin configurations
of altermagnets.

							This paper is organized as follows: In    Sec.~\ref{sec2}, we introduce the model and $T$-matrix formalism. The method for analyzing QPI patterns in the presence of a weak point impurity is detailed in Sec.~\ref{sec3}. In Sec.~\ref{sec4}, we explore the QPI patterns for a pristine $d$-wave altermagnet, as well as for altermagnets with Zeeman splitting or SOC, and present the corresponding numerical results. The findings are summarized in Sec.~\ref{sec5}. Additional technical aspects related to the analysis of QPI patterns are provided in the ~\hyperref[appendixA]{Appendix}.

\section{\label{sec2}Model and T-matrix formalism}

We consider a 2D $d$-wave altermagnet with a Zeeman spin splitting $\Delta$ along the $x$ direction and Rashba SOC 
with strength $\lambda$, and the two-band effective Hamiltonian is written as
\begin{equation}\label{h0}
H_0(\mathbf{k})=\mathbf{k}^2+Jk_xk_y\sigma_z+\Delta\sigma_x+\lambda(k_y\sigma_x-k_x\sigma_y),
\end{equation}
where $\sigma_x$, $\sigma_y$ and $\sigma_z$ are Pauli matrices, and $J$ measures 
the altermagnetic spin splitting. In this paper, we use $\hbar=1$. This Hamiltonian can be diagonalized by introducing

\begin{equation}
U(\mathbf{k})=
\begin{bmatrix}
\cos(\theta_{\mathbf{k}}/2)& ie^{-i\phi_{\mathbf{k}}}\sin(\theta_{\mathbf{k}}/2)\\
ie^{i\phi_{\mathbf{k}}}\sin(\theta_{\mathbf{k}}/2)&\cos(\theta_{\mathbf{k}}/2)
\end{bmatrix},
\end{equation}
where $\tan\theta_{\textbf{k}}=\sqrt{(-\lambda k_x)^2+(\Delta+\lambda k_y)^2}/Jk_xk_y$ and 
$\tan\phi_{\textbf{k}}=(-\lambda k_x)/(\Delta+\lambda k_y)$. Then $H_0(\mathbf{k})=
U(\mathbf{k})\begin{bmatrix}
\omega^+_{\mathbf{k}}&0\\
0&\omega^-_{\mathbf{k}}
\end{bmatrix} U^\dagger(\mathbf{k})$, with eigenvalues

\begin{equation}\label{eq3}
\omega^{\pm}_\mathbf{k}=\mathbf{k}^2\pm\sqrt{(\Delta+\lambda k_y)^2+(-\lambda k_x)^2+(Jk_xk_y)^2}.
\end{equation}

The QPI patterns can be calculated with the standard T-matrix formalism. 
We consider a general potential induced by impurities or defects, 
\begin{equation}
\hat{V}(\mathbf{r})=\sum_\beta V^\beta(\mathbf{r})\sigma_\beta,
\end{equation} 
with $\beta=0,x,y,z$ labeling the scattering channels. Specifically, 
$\beta=0$ corresponds to nonmagnetic impurities and  
$\beta=x,y,z$ denote the directions of local exchange fields induced by magnetic impurities.
In the presence of impurity potentials, interference between 
incident and outgoing waves, \emph{i.e.,} QPI, leads to spacial modulations of the LDOS. 
The spin-resolved LDOS can be extracted by the spin-resolved STS measurements~\cite{Wiesendanger2009,Jeon2017,Cornils2017}, which
is related to the Green's function $G$~\cite{Mitchell2016} as
\begin{equation}
\rho_\alpha(\mathbf{r},\omega)=-\frac{1}{\pi}\Im\{\mathrm{Tr}[\sigma_\alpha G(\mathbf{r},\mathbf{r},\omega)]\},
\end{equation}
where $\alpha=0,x,y,z$ stands for the probe channels of spin~\cite{Guo2010} and $\Im$ denotes the strength of a 
branch cut across the real frequency axis defined by $\Im f(\omega)=[f(\omega+i0^+)-f(\omega-i0^+)]/2i$. 

The Green's function $G(\mathbf{r},\mathbf{r},\omega)=\langle \mathbf{r}|\hat{G}(\omega)|
\mathbf{r}\rangle=\langle \mathbf{r}|\frac{1}{\omega-\hat{H}+i0^+}|\mathbf{r}\rangle$ for 
the full Hamiltionian $\hat{H}=\hat{H}_0+\hat{V}$ has a simple representation in terms of the T-matrix,
\begin{equation}\label{eq6}
\hat{G}(\omega)=\hat{G}_0(\omega)+\hat{G}_0(\omega)\hat{T}(\omega)\hat{G}_0(\omega),
\end{equation}
where $\hat{G}_0(\omega)=\frac{1}{\omega-\hat{H}_0+i0^+}$, and the T-matrix $\hat{T}$ is determined by the Bethe-Salpeter equation as
\begin{equation}\label{eq7}
\hat{T}(\omega)=\hat{V}+\hat{V}\hat{G}_0(\omega)\hat{T}(\omega).
\end{equation}
We denote the $\hat{T}$ from the scattering channel $\beta$ as 
\begin{equation}\label{tbeta}
\hat{T}^\beta(\omega) =\hat{V}^\beta+\hat{V}^\beta\hat{G_0}(\omega)\hat{T}^\beta(\omega),
\end{equation} 
which can be further decomposed as~\cite{Capriotti2003}
\begin{equation}
\hat{T}^\beta=\sum_{\beta'}\hat{T}^{\beta\beta'}\sigma_{\beta'}.
\end{equation} 

The full information of spin-resolved scattering and interference effects 
can be extracted by the FT-LDOS for general probe and scattering channels labeled by $\alpha$ and $\beta$, respectively,
as given by
\begin{equation}\label{eq10}
\delta \rho_{\alpha\beta}(\mathbf{q},\omega)=\int \frac{\mathrm{d}^2 \mathbf{r}}{(2\pi)^2}  e^{-i\mathbf{q}\cdot\mathbf{r}} \Im\mathrm{Tr}\langle\mathbf{r}|\sigma_\alpha\hat{G_0}(\omega)\hat{T}^\beta(\omega)\hat{G_0}(\omega)|\mathbf{r}\rangle.
\end{equation}
Here we have omitted the first term $\hat{G}_0(\omega)$ in Eq.~\eqref{eq6} because it only contributes the density of the states without the impurity at $\mathbf{q}=\mathbf{0}$.
It can be written in the $\mathbf{k}$-space as
\begin{equation}\label{eq11}
\delta \rho_{\alpha\beta}(\mathbf{q},\omega)=-\frac{1}{2\pi i}[\Lambda_{\alpha\beta}(\mathbf{q},\omega)-\Lambda^*_{\alpha\beta}(\mathbf{-q},\omega)],
\end{equation}
where the response function \cite{PEREGBARNEA2005} reads
\begin{equation}\label{eq12}
\Lambda_{\alpha\beta}(\mathbf{q},\omega)=\int\frac{\mathrm{d}^2\mathbf{k}}{(2\pi)^2}\mathrm{Tr}[\sigma_\alpha G_0(\mathbf{k}+\mathbf{q},\omega)T^\beta_{\mathbf{k}+\mathbf{q},\mathbf{k}}(\omega)G_0(\mathbf{k},\omega)].
\end{equation}
$T^\beta_{\mathbf{k}+\mathbf{q},\mathbf{k}}(\omega)=\langle\mathbf{k}+\mathbf{q}|\hat{T}^{\beta}(\omega)
|\mathbf{k}\rangle$ is the Fourier transform of $T^\beta(\mathbf{r},\mathbf{r'},\omega)$, 
and $G_0(\mathbf{k},\omega)=\frac{1}{\omega-H_0(\mathbf{k})+i0^+}$.

In general, the quantity $\delta \rho_{\alpha\beta}(\mathbf{q},\omega)$ is a complex number. The real-space LDOS $\delta \rho_{\alpha\beta}(\mathbf{r},\omega)$ can be decomposed into the symmetric and antisymmetric parts, defined as $\delta\rho^S(\mathbf{r},\omega)=[\delta\rho(\mathbf{r},\omega)+\delta\rho(-\mathbf{r},\omega)]/2$ and $\delta\rho^A(\mathbf{r},\omega)=[\delta\rho(\mathbf{r},\omega)-\delta\rho(-\mathbf{r},\omega)]/2$, respectively, such that $\delta\rho(\mathbf{r},\omega)=\delta\rho^S(\mathbf{r},\omega)+\delta\rho^A(\mathbf{r},\omega)$. Consequently, the real part of FT-LDOS, $\mathrm{Re}[\delta\rho_{\alpha\beta}(\mathbf{q},\omega)]$, corresponds to the Fourier transform of the symmetric LDOS, while the imaginary part, $\mathrm{Im}[\delta \rho_{\alpha\beta}(\mathbf{q},\omega)]$, corresponds to the Fourier transform of the antisymmetric LDOS.  It is convenient to define symmetric response function $\Lambda^S_{\alpha\beta}(\mathbf{q},\omega)=[\Lambda_{\alpha\beta}(\mathbf{q},\omega)+\Lambda_{\alpha\beta}(\mathbf{-q},\omega)]/2$ and antisymmetric response function $\Lambda^A_{\alpha\beta}(\mathbf{q},\omega)=[\Lambda_{\alpha\beta}(\mathbf{q},\omega)-\Lambda_{\alpha\beta}(\mathbf{-q},\omega)]/2$ to write the real and imaginary parts of FT-LDOS explicitly as
\begin{eqnarray}\label{eq13}
\mathrm{Re}[\delta\rho_{\alpha\beta}(\mathbf{q},\omega)]&=&-\frac{1}{\pi}\mathrm{Im} \left[\Lambda^S_{\alpha\beta}(\mathbf{q},\omega)\right], \nonumber    \\
\mathrm{Im}[\delta\rho_{\alpha\beta}(\mathbf{q},\omega)]&=&\frac{1}{\pi}\mathrm{Re} \left[\Lambda^A_{\alpha\beta}(\mathbf{q},\omega)\right].
\end{eqnarray}

\section{\label{sec3} QPI for short-range potential}

Here, we consider a point impurity with short-range potential effectively captured by 
a $\delta$-function as $V(\mathbf{r})=\delta(\mathbf{r})\sum_\beta V_\beta\sigma_{\beta}$, 
whose Fourier transform is $V(\mathbf{k})=\sum_\beta V_\beta\sigma_{\beta}$,
without $\mathbf{k}$ dependence. 
{Note that the impurity is assumed to be located at the origin. Experimentally, this position can be determined from the LDOS pattern, $\rho(r,\omega)$, which exhibits an oscillatory pattern radiating outward from the impurity. As a result, the phase ambiguity in the FT-LDOS expression, Eq.~\eqref{eq10}, can be eliminated.}
The T-matrix for a single scattering channel 
($V(\mathbf{r})=\delta(\mathbf{r})V_\beta\sigma_{\beta}$)
solved by Eq.~\eqref{tbeta} has a simple form,
\begin{equation}\label{eq14}
T^\beta(\omega)=\left[1-V_\beta\sigma_\beta\int\frac{\mathrm{d}^2\mathbf{k}}{(2\pi)^2}G_0(\mathbf{k},\omega)\right]^{-1}V_\beta\sigma_\beta,
\end{equation}
which is independent of $\mathbf{k}$ and $\mathbf{k}'$. 
{Under the Born approximation \cite{Economou2006}}, $\hat{T}$ can be approximated by $\hat{V}$, that is
$T^\beta\simeq V_\beta\sigma_\beta$. In the following, 
we will focus on the regime of single scattering channel,
meaning that different types of impurities do not coexist, and the associated response function reduces to
\begin{equation}\label{eq15}
\Lambda_{\alpha\beta}(\mathbf{q},\omega)=V_\beta\int\frac{\mathrm{d}^2\mathbf{k}}{(2\pi)^2}\mathrm{Tr}
\left[\sigma_\alpha G_0(\mathbf{k}+\mathbf{q},\omega)\sigma_\beta G_0(\mathbf{k},\omega)\right].
\end{equation}
It can be expressed using the Lehmann representation as
\begin{equation}\label{eq16}
\begin{split}
&\Lambda_{\alpha\beta}(\mathbf{q},\omega)=\\
&V_\beta\int\frac{\mathrm{d}^2\mathbf{k}}{(2\pi)^2}\mathrm{Tr}
\left[\sum_{s,s'}\sigma_\alpha \frac{|s,\mathbf{k}+\mathbf{q}\rangle\langle s,
\mathbf{k}+\mathbf{q}|}{\omega-\omega^s_{\mathbf{k}+\mathbf{q}}+i0^+}\sigma_{\beta} 
\frac{|s',\mathbf{k}\rangle \langle s',\mathbf{k}| }{\omega-\omega^{s'}_{\mathbf{k}}+i0^+}\right],
\end{split}
\end{equation}
where $s,s'=\pm1$ denote the two bands, $|s,\mathbf{k}\rangle$ is the eigenstate of 
$H_0(\mathbf{k})$ and $\omega^{s}_{\mathbf{k}}$, $\omega^{s'}_{\mathbf{k}+\mathbf{q}}$ are defined in Eq.~\eqref{eq3}.

Furthermore, we define the spin-coherent factor as~\cite{Zhang2019}
\begin{equation}\label{eq17}
F_{\alpha\beta}^{ss'}=\mathrm{Tr}[\sigma_\alpha|s,\mathbf{k}+\mathbf{q}\rangle\langle s,\mathbf{k}+\mathbf{q}|\sigma_{\beta}|s',\mathbf{k}\rangle\langle s',\mathbf{k}|],
\end{equation}
and decompose the imaginary and real parts of the Green's function as
\begin{eqnarray}\label{eq18}
A_s(\mathbf{k},\omega)&=&\mathrm{Im}\frac{1}{\omega-\omega^{s}_{\mathbf{k}}+i0^+}=-\pi\delta(\omega-\omega^{s}_{\mathbf{k}}), \nonumber    \\
B_s(\mathbf{k},\omega)&=&\mathrm{Re} \frac{1}{\omega-\omega^{s}_{\mathbf{k}}+i0^+}=\mathcal{P}\frac{1}{\omega-\omega^{s}_{\mathbf{k}}}.
\end{eqnarray}
Then the response function can be expressed as
\begin{equation}\label{eq19}
\begin{split}
\Lambda_{\alpha\beta}(\mathbf{q},\omega)&:=V_\beta\tilde{\Lambda}_{\alpha\beta}(\mathbf{q},\omega),\\
\tilde{\Lambda}_{\alpha\beta}(\mathbf{q},\omega)&=\sum_{s,s'}\int\frac{\mathrm{d}^2\mathbf{k}}{(2\pi)^2}F^{ss'}_{\alpha\beta}[A_{s'}(\mathbf{k},\omega)+iB_{s'}(\mathbf{k},\omega)]\\
&\times[A_s(\mathbf{k}+\mathbf{q},\omega)+iB_s(\mathbf{k}+\mathbf{q},\omega)].\\
\end{split}
\end{equation}
Because of the relation between FT-LDOS and the response function Eq.~\eqref{eq11}, 
the analysis of QPI patterns can be divided into the following three steps \cite{Zhang2019} 
(more details can be found in the~\hyperref[appendixA]{Appendix}).

(i) Analyzing the singular behaviors of the response function by ignoring the spin-coherent factor. 
Four possible terms in the response 
function are $A_s(\mathbf{k}+\mathbf{q})A_{s'}(\mathbf{k})$, $-B_s(\mathbf{k}
+\mathbf{q})B_{s'}(\mathbf{k})$, $iB_s(\mathbf{k}+\mathbf{q})A_{s'}(\mathbf{k})$ 
and $iA_s(\mathbf{k}+\mathbf{q})B_{s'}(\mathbf{k})$, which we denote as $AA, BB, BA, AB$ for short. We note that all 
four terms diverge at the same $\mathbf{q}$ point, but in different ways. 
Without loss of generality, we focus on the singular behavior of 
the joint density of state $\int\frac{\mathrm{d}^2\mathbf{k}}
{(2\pi)^2}A_s(\mathbf{k}+\mathbf{q})A_{s'}(\mathbf{k})$,
which gives the conditions of divergent joint density of state as
\begin{eqnarray}\label{eq20}
\omega_{\mathbf{k+q}}^s=\omega, \ \ 
\omega_{\mathbf{k}}^{s'}=\omega, \ \ 
(\nabla\omega_{\mathbf{k+q}}^s)\times(\nabla\omega_{\mathbf{k}}^{s'})=0,
\end{eqnarray}
namely, two points on the isoenergy surfaces
with parallel normal vectors determine a singular scattering vector $\pm\mathbf{q}$
measuring their momentum difference.

(ii) Investigating the effect of the spin-coherent factor $F_{\alpha\beta}^{ss'}$. It is generally a complex number 
which determines how each term (of $AA$-, $AB$-, $BA$- and $BB$-type functions) 
contributes to the real and imaginary parts of FT-LDOS, $\delta\rho_{\alpha\beta}(\mathbf{q},\omega)$. 
Meanwhile, it accounts for further enhancement or suppression of the scattering associated with the spin texture
on the Fermi surface as well as the spin configurations of the impurities.

(iii) Selecting stable singularities by taking into account the effect of finite lifetime. 
For real materials, quasiparticles always possess a finite lifetime,
which indicates that the factor $0^+$ in the Green's function should be replaced by a small positive number $\eta$. 
Only the terms that remain for a realistic $\eta$ are considered to be experimentally relevant.
We argue that the singular behaviors of the
$AA$- and $BB$-type terms are alwayls stable while the singularities of the $AB$- 
and $BA$-type terms are not. Specifically, the divergence 
caused by $AB$- and $BA$-type terms will disappear when the momentum transfer
$\mathbf{q}$ connects the approximately flat or nested isoenergy 
surface as illustrated in Fig.~\hyperref[figA1]{A1(b)}.

Additionally, in the weak potential regime, 
the real and imaginary parts of FT-LDOS can be explicitly interpreted by the spin-coherent factor. 
From Eqs.~\eqref{eq15} and ~\eqref{eq19}, we have
\begin{equation}
\tilde{\Lambda}_{\alpha\beta}(\mathbf{-q})=\tilde{\Lambda}_{\beta\alpha}(\mathbf{q}).
\end{equation}
Combined with Eq.~\eqref{eq13}, we have 
\begin{eqnarray}\label{eq22}
\mathrm{Re}[\delta\rho_{\alpha\beta}(\mathbf{q})]&=&-\frac{1}{\pi}\mathrm{Im} \left[\frac{V_\beta}{2}[\tilde{\Lambda}_{\alpha\beta}(\mathbf{q})+\tilde{\Lambda}_{\beta\alpha}(\mathbf{q})]\right], \nonumber    \\
\mathrm{Im}[\delta\rho_{\alpha\beta}(\mathbf{q})]&=&\frac{1}{\pi}\mathrm{Re} \left[\frac{V_\beta}{2}[\tilde{\Lambda}_{\alpha\beta}(\mathbf{q})-\tilde{\Lambda}_{\beta\alpha}(\mathbf{q})]\right],
\end{eqnarray}
where the independent variable $\omega$ is omitted. The term that replaces the original term $\tilde{\Lambda}_{\alpha\beta}(\mathbf{-q})$ is expressed as
\begin{equation}\label{eq23}
\begin{split}
\tilde{\Lambda}_{\beta\alpha}(\mathbf{q})&=
\sum_{s,s'}\int\frac{\mathrm{d}^2\mathbf{k}}{(2\pi)^2}F^{ss'}_{\beta\alpha}[A_{s'}(\mathbf{k},\omega)+iB_{s'}(\mathbf{k},\omega)]\\
&\times[A_s(\mathbf{k}+\mathbf{q},\omega)+iB_s(\mathbf{k}+\mathbf{q},\omega)].
\end{split}
\end{equation}
Compared to Eq.~\eqref{eq19} with replacing $\mathbf{q}$ by $\mathbf{-q}$, the only difference is that the 
indices $\alpha$ and $\beta$   in the spin-coherent factor $F^{ss'}_{\alpha\beta}$ are swapped. 
It is convenient to define the symmetric (S) and antisymmetric (A) spin-coherent factors as
\begin{eqnarray}\label{eq24}
	F^{ss'}_{S,\alpha \beta}:=\frac{1}{2}(F^{ss'}_{\alpha\beta} + F^{ss'}_{\beta\alpha}), \nonumber    \\
	F^{ss'}_{A,\alpha \beta}:=\frac{1}{2}(F^{ss'}_{\alpha\beta} - F^{ss'}_{\beta\alpha}),
\end{eqnarray}
and the corresponding S(A) reduced response functions as
\begin{equation}
	\begin{split}
		\tilde{\Lambda}^{S(A)}_{\alpha\beta}(\mathbf{q},\omega)&=\sum_{s,s'}\int\frac{\mathrm{d}^2\mathbf{k}}{(2\pi)^2}F^{ss'}_{S(A),\alpha\beta}[A_{s'}(\mathbf{k},\omega)+iB_{s'}(\mathbf{k},\omega)]\\
		&\times[A_s(\mathbf{k}+\mathbf{q},\omega)+iB_s(\mathbf{k}+\mathbf{q},\omega)].
	\end{split}
\end{equation}
Therefore, the real part of the FT-LDOS $\mathrm{Re}[\delta\rho_{\alpha\beta}(\mathbf{q})]=-\frac{1}{\pi}\mathrm{Im} \left[V_\beta\tilde{\Lambda}^S_{\alpha\beta}(\mathbf{q},\omega)\right]$ is 
determined by $F^{ss'}_{S,\alpha \beta}$ with the other terms in Eq.~\eqref{eq19} remaining unchanged, while the imaginary part $\mathrm{Im}[\delta\rho_{\alpha\beta}(\mathbf{q})]=\frac{1}{\pi}\mathrm{Re} \left[V_\beta\tilde{\Lambda}^A_{\alpha\beta}(\mathbf{q},\omega)\right]$ is governed by  $F^{ss'}_{A,\alpha \beta}$.

\begin{figure}[htbp]\label{fig1}
	\centering
	\includegraphics[scale=0.62]{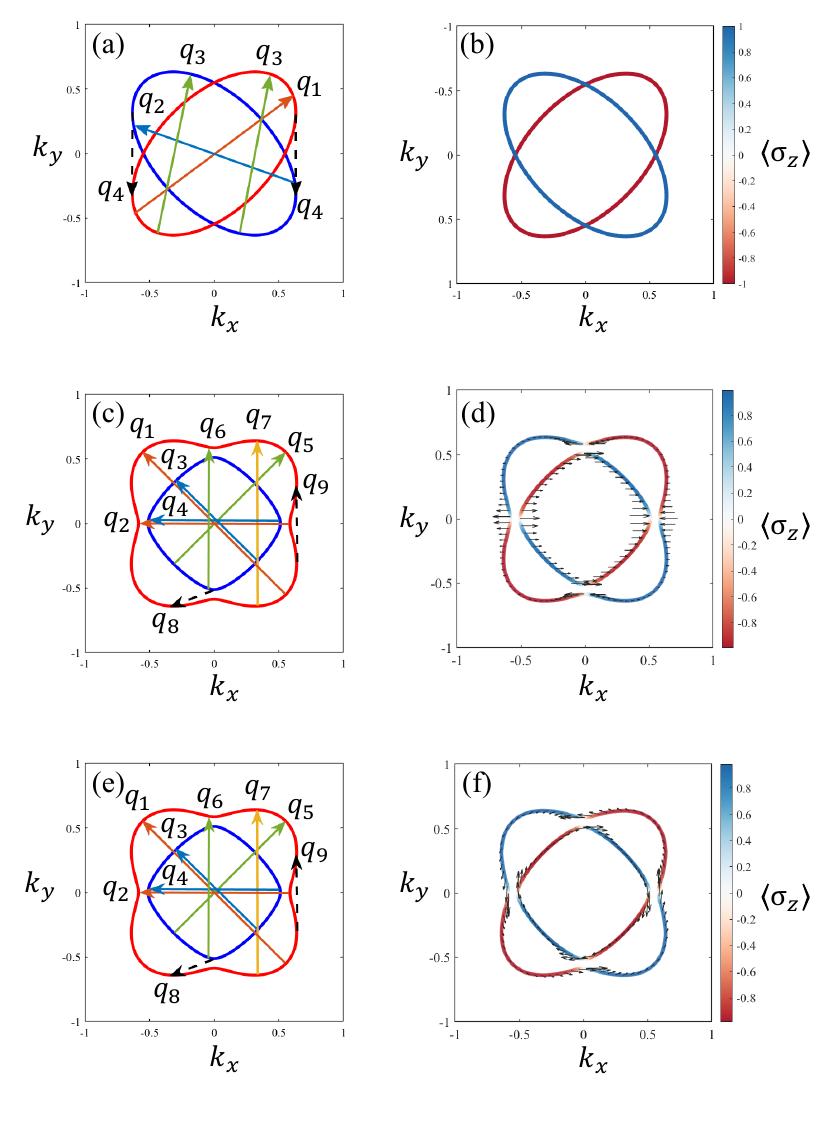}
	\caption{Isoenergy surfaces for various altermagnet cases, with representative divergent scattering vectors and corresponding spin textures. (a) Isoenergy surface of the d-wave altermagnet, with scattering vectors $\mathbf{q}_1$ to $\mathbf{q}_4$, and (b) the associated spin texture at $\omega = 0.3$, $J = 1$.
		For the altermagnet with Zeeman splitting, (\textcolor{red}{c}) shows the isoenergy surface with scattering vectors $\mathbf{q}_1$ to $\mathbf{q}_9$, and (\textcolor{red}{d}) depicts the spin texture at $\omega = 0.3$, $J = 1$, and $\Delta = 0.1$.
		For the altermagnet with spin-orbit coupling (SOC), the isoenergy surface and scattering vectors $\mathbf{q}_1$ to $\mathbf{q}_9$ are shown in (\textcolor{red}{e}), while the spin texture for $\omega = 0.3$, $J = 1$, and $\lambda = 0.1$ is presented in (\textcolor{red}{f}).
		Dashed arrows indicate unstable scattering vectors for quasiparticles with finite lifetimes.}
\end{figure}

\begin{figure*}[htbp]\label{fig2}
	\centering
	\includegraphics[scale=0.55]{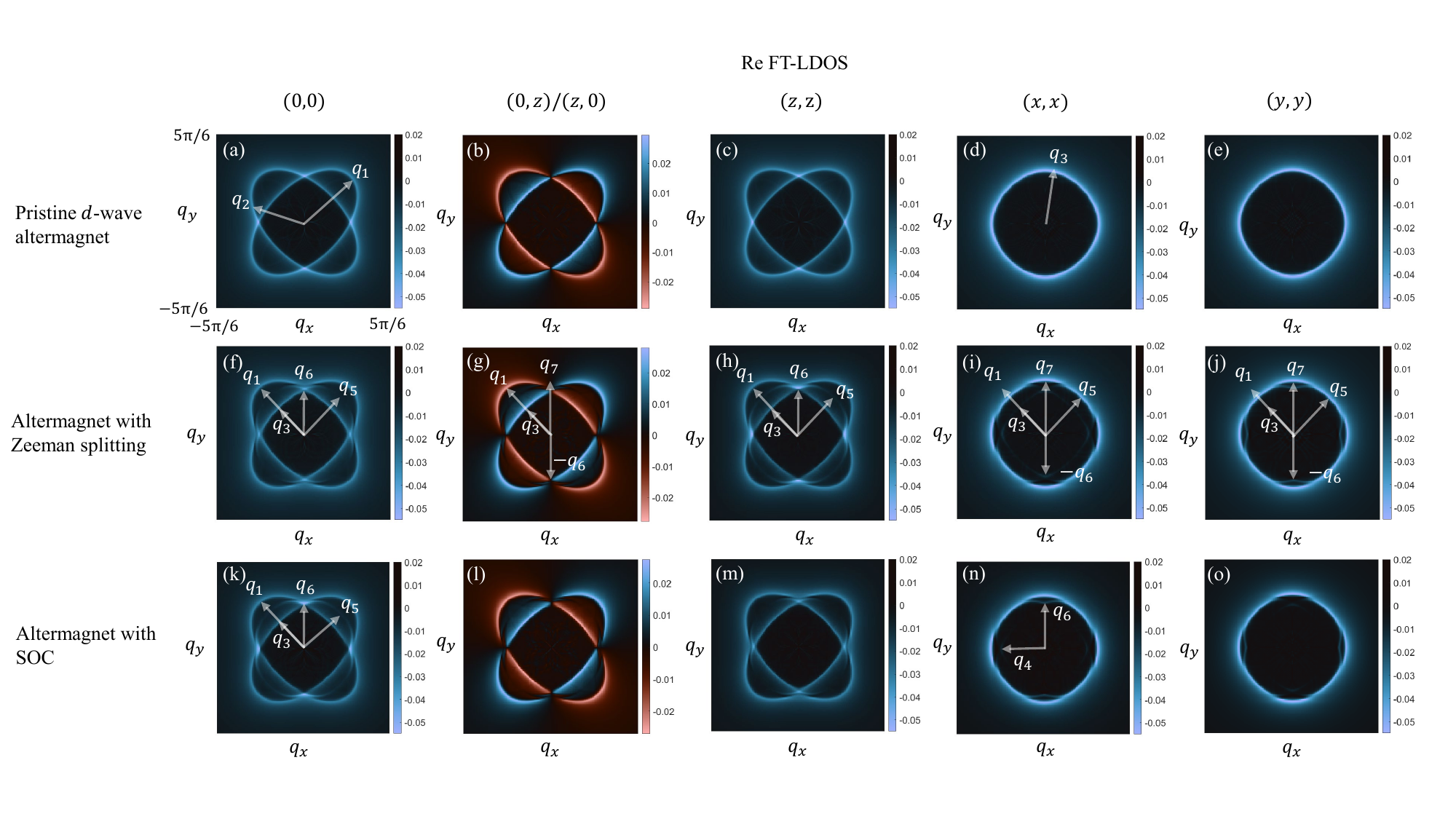}
	\caption{The real parts of FT-LDOS (Re FT-LDOS) are shown for 
		different altermagnet cases and channels. Representative scatterings are 
		depicted in panels (a), (c), (f), (h), (k) and (m). Parameters are set to $\omega = 0.3$, $J = 1$ for pristine $d$-wave altermagnet; $\omega = 0.3$, $J = 1$, $\Delta=0.04$ for altermagnet with Zeeman splitting; $\omega = 0.3$, $J = 1$, $\lambda=0.075$ for altermagnet with SOC, and $V=V_0\sigma_\beta$, $\eta=0.013$ and $V_0=0.05$.}
\end{figure*}

\section{\label{sec4} QPI in altermagnets}

\subsection{Pristine $d$-wave altermagnet}

For a pristine $d$-wave altermagnet, we have $\Delta=\lambda=0$ for $H_0$ in Eq.~\eqref{h0}.
The isoenergy surfaces for $\omega=0.3$, $J=1$ and the spin polarization are shown 
in Figs.~\hyperref[fig1]{1(a)} and \hyperref[fig1]{1(b)}, respectively,
which exhibit altermagnetic spin splitting of $d$-wave symmetry, 
\emph{i.e.}, two ellipses with orthogonal major axes and opposite spin polarization
in the $z$ direction. According to Eq.~\eqref{eq20}, 
there are singular scattering vectors $\mathbf{q}$ determined by the normal vectors 
on the isoenergy surface, and they can be further divided into intra-band (same spin) and inter-band (opposite spin) ones. 
Different categories of singular scattering vectors $\mathbf{q}_1$, $\mathbf{q}_2$, $\mathbf{q}_3$ and
$\mathbf{q}_4$ are illustrated in Fig.~\hyperref[fig1]{1(a)}, where $\mathbf{q}_1$ 
and $\mathbf{q}_2$ denote the intra-band scattering, 
$\mathbf{q}_3$, and $\mathbf{q}_4$ correspond to inter-band scattering. Due to the
$d$-wave altermagnetic symmetry of the isoenergy surfaces, 
the intra-band scattering takes place between $\mathbf{k}$ and $\mathbf{-k}$, 
which is the backscattering. Meanwhile, the inter-band scattering is repeated twice, from the 
spin-up band to the spin-down band or vice versa, as illustrated by the double arrows for $\mathbf{q}_3$ and $\mathbf{q}_4$
in Fig.~\hyperref[fig1]{1(a)}.
Note that the signature due to the scattering with a momentum transfer $\mathbf{q_4}$
is unstable for a finite lifetime as discussed in the~\hyperref[appendixA]{Appendix},
which will not be reflected in the FT-LDOS patterns. 

\begin{table}[!ht]
	\centering
	\label{table1}
	\caption{Spin-coherent factors $F_{\alpha\beta}^{ss'}$}
	\begin{tabular}{ccccc}
		\hline
		$\alpha\backslash\beta$& $0$ & $x$ & $y$ & $z$ \\ \hline
		$0$ & $\frac{1}{2}(1+ss')$ & $0$ & $0$ & $\frac{1}{2}(s+s')$ \\
		$x$ & $0$ & $\frac{1}{2}(1-ss')$ & $\frac{i}{2}(s'-s)$ & $0$ \\
		$y$ & $0$ & $\frac{i}{2}(s-s')$ & $\frac{1}{2}(1-ss')$ & $0$ \\
		$z$ & $\frac{1}{2}(s+s')$ & $0$ & $0$ & $\frac{1}{2}(1+ss')$ \\ \hline
	\end{tabular}
\end{table}

For pristine altermagnets,
the spin-coherent factor $F_{\alpha\beta}^{ss'}$ reduces to a simple form as the spin texture is independent 
of the momenta $\mathbf{k}$ and $\mathbf{k+q}$. For various probe and scattering channels denoted by the ordered labels 
$(\alpha,\beta)$, the spin-coherent factors are 
summarized in TABLE~\hyperref[table1]{\uppercase\expandafter{\romannumeral1}}, which are calculated by Eq.~\eqref{eqA9}.
%Next, we denote the probe and scattering channels as ordered labels 
%$(\alpha,\beta)$ and sum over $s$ and $s'$ to obtain the response function given in 
%Eq.~\hyperref[eq16]{(16)}, which determines the QPI pattern. 
Specifically, $\alpha$
corresponds to the spin polarization of the STS measurements and $\beta$ is the spin
direction of the magnetic impurity.
From TABLE~\hyperref[table1]{\uppercase\expandafter{\romannumeral1}}, one can see that the spin-coherent factors depend only on the band indices $s$ and $s'$.
It is evident that there is no response in 
the channels $(0,x), (0,y), (x,0), (x,z), (y,0), (y,z), (z,x)$, 
and $(z,y)$. Channels $(x,y)$ and $(y,x)$ do not contribute to the FT-LDOS either despite the nonzero $F_{\alpha\beta}^{ss'}$ factors.
First, the real part of the FT-LDOS is zero because $F^{ss'}_{xy}=-F^{ss'}_{yx}$ and so the corresponding symmetric 
spin-coherent factor $F^{ss'}_{S,xy}$ defined by Eq.~\eqref{eq24} vanishes.
Second, there is no response in its imaginary part dominated by the antisymmetric 
spin-coherent factor $F^{ss'}_{A,xy}$ because either the intra-band scattering 
($\mathbf{q}_1$ and $\mathbf{q}_2$ in Fig.~\hyperref[fig1]{1(a)}) is forbidden by $s-s'=0$, or
the contributions by the inter-band scattering ($\mathbf{q}_3$ and $\mathbf{q}_4$ in 
Fig.~\hyperref[fig1]{1(a)}) cancel each other out
after summing over $s$ and $s'$. 

For the remaining channels, TABLE \hyperref[table1]{\uppercase\expandafter{\romannumeral1}} further indicates that only the 
real part of the FT-LDOS is present, as their antisymmetric spin-coherent factors $F^{ss'}_{A,\alpha \beta}$ equal zero. 
According to Eqs.~\eqref{eq22} and~\eqref{eq23}, only $AB$- and $BA$-type terms
contribute to the real part of the FT-LDOS, which may become unstable when a finite lifetime $\eta$ is taken into account. 
From the spin-coherent factors, the responses in channels 
$(0,0), (0,z), (z,0)$, and $(z,z)$ are nonzero only when $s = s'$, indicating an intra-band backscattering. 
This type of scattering between $\mathbf{k}$ and $\mathbf{-k}$ on the isoenergy surface allows the momentum transfer 
$\mathbf{q}_{1,2}=2\textbf{k}$ to reproduce the isoenergy configurations in the FT-LDOS with a
doubled scale as shown in Figs.~\hyperref[fig2]{2(a,b,c)}.
In particular, the spin-coherent factors of $(0,z)$ and $(z,0)$ have the form of $s+s'$, which yield FT-LDOS patterns 
with opposite signs for different bands, as shown in Fig.~\hyperref[fig2]{2(b)}.
Therefore, the FT-LDOS measurements in the ($0,z$) and ($z,0$) channels clearly reveal the pristine
altermagnetic Fermi surfaces.
The factors for channels $(x,x)$ and $(y,y)$ are nonzero only when $s = -s'$, providing geometric 
information about the isoenergy surfaces through inter-band scatterings denoted by $\mathbf{q}_3$.
This also indicates the presence of opposite $z$-direction spins in the two bands.
In the above discussions, the FT-LDOS $\delta \rho_{\alpha,\beta} (\mathbf{q},\omega)$ for different 
channels are numerically calculated by Eq.~\eqref{eq11}. 

\subsection{Altermagnet with Zeeman splitting}

\begin{figure}[htbp]\label{fig3}
	\centering
	\includegraphics[scale=0.73]{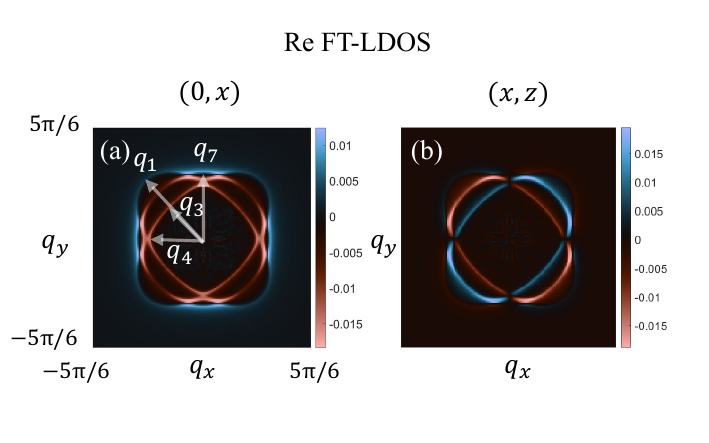}
	\caption{The real parts of FT-LDOS (Re FT-LDOS) of altermagnet with Zeeman splitting are shown for additional channels. Parameters are set to $\omega = 0.3$, $J = 1$, $\Delta =0.04$,  $V=V_0\sigma_\beta$, $\eta=0.013$ and $V_0=0.05$.}
\end{figure}

The distinct spin configurations of the pristine altermagnetic Fermi surfaces 
imply a unique response when driven by spin-related physical effects.
For the altermagnet with a Zeeman splitting along the $x$ direction, we consider $\lambda=0$, $J=1$ and $\Delta=0.04$ for $H_0$ in Eq.~\eqref{h0}.
The isoenergy surface for $\omega=0.3$ and the spin configurations are shown 
in Figs.~\hyperref[fig1]{1(c)} and \hyperref[fig1]{1(d)}, respectively.
We observe that the in-plane Zeeman field lifts the spin degeneracy along the symmetric lines $k_x=0$ and $k_y=0$, 
triggering a Lifshitz transition of the $d$-wave altermagnetic Fermi surfaces. 
This transition alters the isoenergy surfaces from an elliptical to a petal-like shape, 
accompanied by more complex spin configurations~\cite{Li2024}. 
{In contrast, conventional metals with Zeeman splitting exhibit circular Fermi surfaces, where the spin directions on the inner and outer rings are completely opposite. The corresponding QPI pattern will be isotropic.}
Analyzing the normal vectors on the isoenergy surfaces, 
we identify stable divergent scattering vectors, including intra-lower-band (outer contour) scatterings like $\mathbf{q}_1$, 
$\mathbf{q}_2$ and $\mathbf{q}_7$, intra-upper-band (inner contour) backscatterings such as $\mathbf{q}_3$ and $\mathbf{q}_4$, 
and inter-band scatterings like $\mathbf{q}_5$ and $\mathbf{q}_6$; see Fig.~\hyperref[fig1]{1(c)}. Similarly,
the scattering processes denoted by $\mathbf{q}_8$ and $\mathbf{q}_9$ correspond to unstable divergence,
which may have no contribution to the FT-LDOS. These vectors serve as representatives 
to capture the characteristics of the isoenergy surfaces and help us explore the effects of spin-coherent 
factors in different channels. 
The Zeeman splitting $\Delta$ induces a spin deflection 
towards the $x$ axis by an angle of $\theta_{\textbf{k}}=\tan^{-1} (\Delta/Jk_xk_y)$ from 
the original $\pm z$ directions for the lower and upper bands, respectively. The spin-coherent factors of different channels
that modify the FT-LDOS depend on the spin polarizations on the isoenergy surfaces shown in Fig.~\hyperref[fig1]{1(d)}.

%The backscatterings $\mathbf{q_1}$, $\mathbf{q_2}$, $\mathbf{q_3}$ and $\mathbf{q_4}$ still link two points with full spin overlap. However, the directions of spin at the ends of these backscatterings are different, which are shown in Fig.~\hyperref[fig3]{3(b)}. Intra-band scattering $\mathbf{q_7}$ is accompanied by a complete spin overlap if the $z$-spin of one of the endpoints is flipped. The inter-band scattering $\mathbf{q_6}$ have . Scattering $\mathbf{q_5}$ have a small spin overlap. 

%Backscatterings link two points with full spin overlap, and lower-band 
%backscattering $\mathbf{q_1}$ exhibit a small spin overlap along the $x$ direction, while upper-band 
%backscattering $\mathbf{q_3}$ shows a small overlap in the $z$ direction. Conversely, backscatterings 
%$\mathbf{q_2}$ and $\mathbf{q_4}$ feature nearly full spin overlap in the $x$ direction and almost no 
%overlap in the $z$ direction. Inter-band scattering $\mathbf{q_5}$ displays spin overlaps in both the 
%$x$ and $z$ directions. Scattering $\mathbf{q_6}$ have only large $z$-direction spin overlap while 
%$\mathbf{q_7}$ have only small $x$-direction overlap.

For the channel $(0,y)$, the symmetric spin-coherent factor is zero, $F^{ss'}_{S,0y}=0$, and the 
antisymmetric factor 
\begin{equation}
	F^{ss'}_{A,0y}=\frac{1}{2}iss'\frac{\Delta(Jk_xk_y-Jk'_x k'_y)}{\sqrt{(\Delta^2+J^2k_x^2k_y^2)
(\Delta^2+J^2 {k'_x}^2 {k'_y}^2 )}},
\end{equation} 
with $\mathbf{k}'=\mathbf{k}+\mathbf{q}$. 
The backscattering $\mathbf{k}\leftrightarrow -\mathbf{k}$ is forbidden because $Jk_xk_y-Jk'_x k'_y=0$ and the inter-band scatterings 
cancel each other out by summation over $s$ and $s'$. The same conclusion applies to the $(x,y)$ and $(y,z)$
channels. Therefore, the whole FT-LDOS patterns, including both real and imaginary parts, 
are zero in the $(0,y)$, $(x,y)$ and 
$(y,z)$ channels. 
For the remaining channels, it is straightforward to 
verify that the antisymmetric spin-coherent factors always vanish with $F^{ss'}_{A,\alpha\beta}=0$.
%Specifically, the factors in diagonal channels are zero by definition, meaning $F^{ss'}_{A,\alpha\alpha}=0$.
As a result, only the real parts of the FT-LDOS remain, which are determined by $F^{ss'}_{S,\alpha\beta}=F^{ss'}_{\alpha\beta}$.

We numerically calculated the real parts of the FT-LDOS in various channels and
present part of the results in Figs.~\hyperref[fig2]{2(f-j)} for comparison with other scenarios.
The remaining results are plotted in Fig.~\hyperref[fig3]{3} which is absent for the pristine altermagnets.
The Hamiltonian in Eq.~\eqref{h0} can be written as $H_0(\mathbf{k})=\mathbf{k}^2+\mathbf{d}_{\mathbf{k}}\cdot\boldsymbol{\sigma}$, with $\mathbf{d}_{\mathbf{k}}=(\mathbf{d}_{\mathbf{k}}^x, \mathbf{d}_{\mathbf{k}}^y, \mathbf{d}_{\mathbf{k}}^z)$ and $\boldsymbol{\sigma}=(\sigma_x,\sigma_y,\sigma_z)$.
Then the symmetric spin-coherent factor of channel $(0,0)$ possesses the form of
\begin{equation}\label{eq27}
	F^{ss'}_{S,00}=\frac{1}{2}(1+ss'\hat{\mathbf{d}}_{\mathbf{k}+\mathbf{q}}\cdot \hat{\mathbf{d}}_{\mathbf{k}}),
\end{equation}
where $\hat{\mathbf{d}}_{\mathbf{k}}=[\sum_i (\mathbf{d}_{\mathbf{k}}^i )^2]^{-\frac{1}{2}}(\mathbf{d}_{\mathbf{k}}^x,\mathbf{d}_{\mathbf{k}}^y,\mathbf{d}_{\mathbf{k}}^z)$ has been normalized.
The above equation indicate that $F^{ss'}_{S,00}$ is determined by the spin overlap between $\mathbf{k}+\mathbf{q}$ and $\mathbf{k}$ states. Therefore, 
the corresponding FT-LDOS pattern in Fig.~\hyperref[fig2]{2(f)}
shows maximum intensities induced by the backscatterings because of perfect spin alignment of the 
scattering states, $F^{ss'}_{S,00}=1$, denoted by $\mathbf{q}_1$ ($\mathbf{q}_2$) and $\mathbf{q}_3$ ($\mathbf{q}_4$).
These features inherit from the pristine altermagnet in Fig.~\hyperref[fig2]{2(a)} apart from
the modifications occuring around the band degenerate regions.
The FT-LDOS for $\mathbf{q}_5$ that is absent in pristine altermagnet
is still not visible because a small Zeeman field is insufficient to
align the originally opposite spins, \emph{i.e.}, $F^{00}_{S,\alpha\beta}(\mathbf{q}_5)\sim0$.
However, new intensities in the vicinity of $\mathbf{q}_6$ arise due to inter-band scattering with finite $F^{ss'}_{S,00}$
[cf. Fig.~\hyperref[fig1]{1(c)}]. They stem from the gap opening along the originally degenerate lines in momentum space and the associated Lifshitz transition of the Fermi surfaces induced by the Zeeman field. This effect highlights the deformation of altermagnetic Fermi surfaces under the influence of the Zeeman field, providing direct evidence for the identification of altermagnetism.

For the $(0,z)$ channel, the factor is written as 
\begin{equation}\label{eq28}
	F^{ss'}_{S,0z}=\frac{1}{2}[s\frac{Jk'_xk'_y}{\sqrt{
			(\Delta^2+J^2 {k'_x}^2 {k'_y}^2 )}}+ s'\frac{Jk_xk_y}{\sqrt{(\Delta^2+J^2k_x^2k_y^2)}}],
\end{equation}
and the corresponding pattern is shown in Fig.~\hyperref[fig2]{2(g)}. Considering backscattering 
satisfying $J{k'}_x{k'}_y=J{k}_x{k}_y$, the sign of $F^{ss'}_{S,0z}$ depends on 
which quadrant $\mathbf{q}$ is located in. It is also related to the $z$-component of the spin texture according to Eq.~\eqref{eqA11}.
For examples, $\mathbf{q}_1$ links two states which both have $\langle\sigma_z\rangle>0$, inducing negative intensity at $\mathbf{q}_1$, $F^{ss'}_{S,0z}(\mathbf{q}_1)<0$. In contrast, we have positive intensity at $\mathbf{q}_3$, $F^{ss'}_{S,0z}(\mathbf{q}_3)>0$. The intensities induced by inter-band scattering $\mathbf{q}_5$ are canceled after summing over $s$ and $s'$, and $\mathbf{q}_7$ is directly suppressed by $sJk'_xk'_y+s'Jk_xk_y=0$. The ($0,z$) pattern  in Fig.~\hyperref[fig2]{2(g)} inherits the main features of pristine altermagnet in Fig.~\hyperref[fig2]{2(b)} except for a few slight deformations in shape and newly emerging intensities marked by $\mathbf{q}_6$.

The spin-coherent factors for diagonal channels $(\alpha,\alpha)$, 
\begin{equation}\label{eq29}
	F_{\alpha\alpha}^{ss'}=\frac{1}{2}(1+ss'\hat{\mathbf{d}}_{\mathbf{k}+\mathbf{q}}\cdot \hat{R}_\alpha\hat{\mathbf{d}}_{\mathbf{k}}),
\end{equation}
with $\hat{R}_\alpha$ being a $\pi$-rotation about the $\alpha (=x,y,z)$ axis,
measure the overlap between the spin of the $\mathbf{k}+\mathbf{q}$ state and that of the $\mathbf{k}$ state with a flipping. 
It indicates that the enhancement or suppression of the QPI signatures
strongly depends on the spin directions of the scattering states. 
For the ($z,z$) channel, the spin-coherent factor of the inter-band scattering denoted by $\mathbf{q}_5$($\mathbf{q}_6$) is increased by the spin rotation $\hat{R}_z$, which inverts the $x$-component of spin in the $\mathbf{k}$ state and so
enhances the its overlap with the spin in the $\mathbf{k+q}$ state. Therefore, 
compared to Fig.~\hyperref[fig2]{2(f)}, ($z,z$) pattern in Fig.~\hyperref[fig2]{2(h)} have been slightly enhanced at $\mathbf{q}_5$($\mathbf{q}_6$).
In contrast, the spin flipping reduces the spin overlap between scattering states denoted by $\mathbf{q}_1$($\mathbf{q}_2$) and $\mathbf{q}_3$($\mathbf{q}_4$),
and correspondingly, the QPI patterns denoted by these momenta are slightly suppressed.
Similarly, in the pattern of ($x,x$) channel shown in Fig.~\hyperref[fig2]{2(i)}, backscattering on the lower band $\mathbf{q}_1$($\mathbf{q}_2$) 
is strongly suppressed by $F^{ss'}_{S,xx}\sim0$, while $\mathbf{q}_5$ and $\mathbf{q_7}$ are enhanced by $F^{ss'}_{S,xx}\sim1$. These features are 
inherited from pristine altermagnet in Fig.~\hyperref[fig2]{2(d)}. In addition, new features with weak intensity
arise due to the upper-band backscattering denoted by $\mathbf{q}_3$($\mathbf{q}_4$) and the inter-band scattering marked by $\mathbf{q}_6$, 
both stemming from the nonzero $x$-directional spin induced by 
the Zeeman field. The ($y,y$) pattern in Fig.~\hyperref[fig2]{2(j)} also inherits the features from pristine altermagnet apart from stronger intensities at $\mathbf{q}_6$ and $\mathbf{q}_7$. 
Different from ($x,x$) pattern in 
Fig.~\hyperref[fig2]{2(i)}, the upper-band backscattering denoted by $\mathbf{q}_3$($\mathbf{q}_4$) is completely suppressed with 
$F^{00}_{S,yy}\sim0$, and meanwhile, the pattern marked by $\mathbf{q}_6$ is weakly enhanced.

Patterns of ($0,x$) and ($x,z$) channels shown in Fig.~\hyperref[fig3]{3} are absent for the pristine altermagnets. The spin-coherent 
factor for the ($0,x$) channel reads
\begin{equation}\label{eq30}
	F^{ss'}_{S,0x}=\frac{1}{2}[s\frac{\Delta}{\sqrt{
			\Delta^2+J^2 {k'_x}^2 {k'_y}^2 }}+ s'\frac{\Delta}{\sqrt{\Delta^2+J^2k_x^2k_y^2}}],
\end{equation}
which explains that the pattern in Fig.~\hyperref[fig3]{3(a)} has 
positive and negative values corresponding to the lower-band ($\mathbf{q}_{1,7}$)
and upper-band ($\mathbf{q}_{3,4}$) scatterings, respectively. The factor $F^{ss'}_{S,0x}$ is finite for all inter-band 
scatterings, which help us recognize the closed loop in the pattern. Furthermore, $F^{ss'}_{S,0x}$ attains larger values at scattering wave vectors $\mathbf{q}$ with smaller module $|\mathbf{q}|$, characterized by $(\Delta^2+J^2k_x^2k_y^2)(\Delta^2+J^2 {k'_x}^2 {k'_y}^2 )$. This results in a modulation of intensity as $\mathbf{q}$ varies, for instance, the intensity at $\mathbf{q}_7$ is stronger than which at $\mathbf{q}_1$. 
The spin-coherent 
factor for the ($x,z$) channel is expressed as
\begin{equation}\label{eq31}
	F^{ss'}_{S,xz}=\frac{1}{2}ss'\frac{\Delta (J k_x k_y+J {k'}_x {k'}_y)}{\sqrt{\Delta^2+J^2k_x^2k_y^2}
			\sqrt{\Delta^2+J^2 {k'_x}^2 {k'_y}^2 }},
\end{equation}
which modifies the pattern intensities mainly through the module 
$(\Delta^2+J^2k_x^2k_y^2)(\Delta^2+J^2 {k'_x}^2 {k'_y}^2)$. Such that the lower-band (outer isoenergy contour)
and upper-band (inner isoenergy contour)
scattering has weaker and stronger intensities, respectively; see Fig.~\hyperref[fig3]{3(b)}. 
The intensities corresponding to inter-band scattering are most visible because of the double counting
when summing over $s$ and $s'$.

We have observed that the Zeeman field can introduce distinctive modifications to the FT-LDOS of altermagnets, 
offering valuable information for their identification. Specifically,
the Lifshitz transition of the 
Fermi surfaces can be revealed through the deformation of the FT-LDOS patterns in the
backscattering channels. Zeeman splitting further alters the spin texture, opening additional scattering 
channels and thereby enriching the structures of the FT-LDOS in a distinctive manner.

\subsection{Altermagnet with Rashba SOC}

\begin{figure}[htbp]\label{fig4}
	\centering
	\includegraphics[scale=0.57]{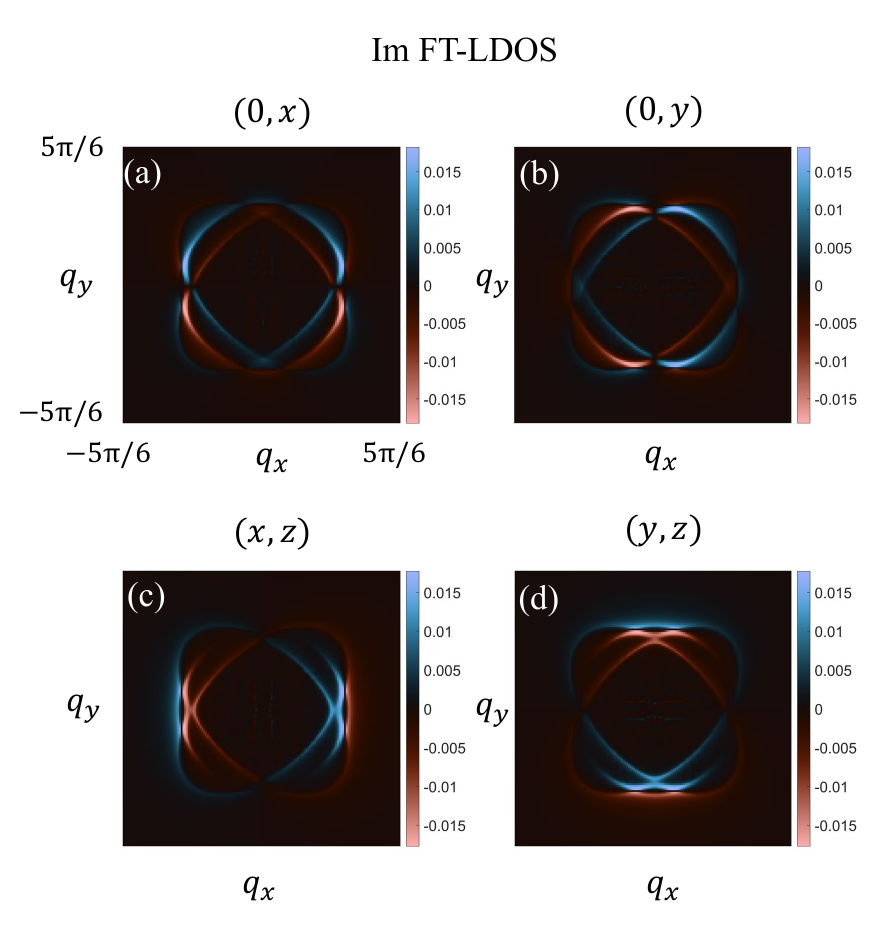}
	\caption{The imaginary parts of FT-LDOS (Im FT-LDOS) of altermagnet with Zeeman splitting are shown for additional channels. Parameters are set to $\omega = 0.3$, $J = 1$, $\lambda =0.075$,  $V=V_0\sigma_\beta$, $\eta=0.013$ and $V_0=0.05$.}
\end{figure}

In real material candidates for altermagnets, the Rashba SOC often exists\cite{Dieny2017,Barnes2014}.
We consider $\lambda=0.075$, $J=1$ and $\Delta=0$ for $H_0$ in Eq.~\eqref{h0}.
The isoenergy surfaces for $\omega=0.3$ shown in Fig.~\hyperref[fig1]{1(e)} is identical to 
those in Fig.~\hyperref[fig1]{1(c)} with a Zeeman field, though their spin textures are quite different;
compare Figs.~\hyperref[fig1]{1(f)} and ~\hyperref[fig1]{1(d)}.
In particular, the Rashba SOC induces a nontrivial spin winding along both isoenergy contours,
in contrast to the situation under the Zeeman field. 
Quantitatively, the Rashba SOC induces a spin deflection 
towards the $x$ direction by an angle of $\theta_{\textbf{k}}=\tan^{-1}[\sqrt{(-\lambda k_x)^2+(\lambda k_y)^2}/Jk_xk_y]$, 
followed by another deflection towards the $y$ direction by $\phi_{\textbf{k}}=\tan^{-1}[(-\lambda k_x)/(\lambda k_y)]$ from 
the original $\pm z$ directions for the lower and upper bands, respectively.
The spin-coherent factors of different channels that dominate the QPI patterns
depend on the spin textures shown in Fig.~\hyperref[fig1]{1(f)}.
{Therefore, altermagnets with SOC exhibit more complex geometries and spin textures compared to conventional metals with Rashba SOC.  In conventional metals, the Fermi surfaces are circular, with spin winding either clockwise or counterclockwise on the inner and outer rings, giving rise to isotropic QPI patterns.}
We consider the representative diverging scattering vectors $\mathbf{q}_1$ to $\mathbf{q}_9$ and study the FT-LDOS
in a parallel way to the regime with a Zeeman field. The numerical results are shown in Figs.~\hyperref[fig2]{2(k-o)} 
and Fig.~\hyperref[fig4]{4}. 
In Figs.~\hyperref[fig2]{2(k-o)}, differences in the in-plane spin components for the SOC and Zeeman field scenarios lead to delicate difference in the QPI patterns. Meanwhile, the patterns of additional channels in Fig.~\hyperref[fig4]{4} have more unique features.

For the channels $(0,0)$, $(x,x)$, $(y,y)$, $(z,z)$ and $(0,z)$, the antisymmetric spin-coherent factor vanishes, \emph{i.e.},
$F^{ss'}_{A,\alpha \beta}=0$, and therefore the symmetric factor $F^{ss'}_{\alpha \beta}=F^{ss'}_{S,\alpha \beta}$,
giving rise to real FT-LDOS $\delta\rho_{\alpha\beta}(\mathbf{q})$. 
The differences between the FT-LDOS patterns in the Zeeman and SOC scenarios
stem from differenct spin modulations, which are strongest around the gap-opening regions.
For the ($0,0$) channel, the spin-coherent factor $F^{ss'}_{S,00}$ (given by Eq.~\eqref{eq27}) that measures the spin overlap between
the states related by $\mathbf{q}_{5,6}$ is larger in the SOC scenario due to the preservation
of time-reversal symmetry. This is reflected by the enhancement of the FT-LDOS patterns denoted by $\mathbf{q}_{5,6}$ in Fig.~\hyperref[fig2]{2(k)} compared with
that in Fig.~\hyperref[fig2]{2(f)}. 

For other diagonal channels ($\alpha,\alpha$), where $\alpha=x,y,z$, 
the spin-coherent factors $F_{\alpha\alpha}^{ss'}$ are given by Eq.~\eqref{eq29}. 
A similar analysis can be conducted, focusing on the differences in the 
in-plane spin components between the two scenarios. The ($z,z$) 
pattern of SOC scenario is shown in Fig.~\hyperref[fig2]{2(m)}, which has the same features as the ($0,0$) pattern in the 
Zeeman-field scenario in Fig.~\hyperref[fig2]{2(f)}. Compared to the ($z,z$) pattern of Zeeman-field scenario in Fig.~\hyperref[fig2]{2(h)}, the inter-band 
scattering is weakly suppressed while the intra-band scattering is enhanced. 
The ($x,x$) pattern in Fig.~\hyperref[fig2]{2(n)} inherits the features of the pristine altermagnet in Fig.~\hyperref[fig2]{2(d)} apart from weak intensities that arise near $\mathbf{q}_4$ and $\mathbf{q}_6$. 
The ($y,y$) pattern in Fig.~\hyperref[fig2]{2(o)} differs from that of the ($x,x$) channel by a rotation of $\pi/2$.

For the ($0,z$) channel, the spin-coherent factor is given by
\begin{equation}
	F^{ss'}_{S,0z}=\frac{1}{2}[s\frac{Jk'_xk'_y}{\sqrt{\lambda^2{k'}^2+J^2{k'_x}^2{k'_y}^2}}+s'\frac{J{k}_x{k}_y}{\sqrt{\lambda^2{k}^2+J^2{k}_x^2{k}_y^2}}],
\end{equation}	
which is similar to Eq.~\eqref{eq28}, with $\Delta$ replaced by $\lambda k^2$ and $\lambda k'^2$. Therefore, the $(0,z)$ pattern of altermagnet with SOC in Fig.~\hyperref[fig2]{2(l)} exhibits the same feature as the one in Fig.~\hyperref[fig2]{2(g)} for the Zeeman-field scenario.

For the channel ($0,x$), the symmetric spin-coherent factor $F^{ss'}_{S,0x}$ has the form
\begin{equation}
	F^{ss'}_{S,0x}=\frac{1}{2}[s\frac{\lambda k'_y}{\sqrt{\lambda^2{k'}^2+J^2{k'_x}^2{k'_y}^2}}+s'\frac{\lambda k_y}{\sqrt{\lambda^2k^2+J^2k_x^2k_y^2}}],
\end{equation}
while the antisymmetric factor $F^{ss'}_{A,0x}$ is given by
\begin{equation}\label{eq33}
	F^{ss'}_{A,0x}=\frac{i}{2}ss'\frac{\lambda J(-k'_xk_xk_y+k_xk'_xk'_y)}{\sqrt{\lambda^2k^2+J^2k_x^2k_y^2}\sqrt{\lambda^2{k'}^2+J^2{k'_x}^2{k'_y}^2}}.
\end{equation}
The symmetric factor $F^{ss'}_{S,0x}$ has no contribution to the real part of FT-LDOS, because it is zero for the dominant intra-band backscattering ($\mathbf{k}\leftrightarrow\mathbf{-k}$) and the contribution due to the inter-band scattering vanishes after summing over $s$ and $s'$. Meanwhile, the imaginary part of FT-LDOS determined by $F^{ss'}_{A,0x}$, shown in Fig.~\hyperref[fig4]{4(a)}, is nonzero. 
$F^{ss'}_{A,0x}$ in Eq.~\eqref{eq33} has the similar form with $F^{ss'}_{S,xz}$ of the Zeeman splitting scenario in Eq.~\eqref{eq31}. A notable correspondence arises in the numerator, where $\Delta$ is replaced by either $-\lambda k'_x$ or $\lambda k_x$.
Therefore, the ($0,x$) pattern exhibits strongest intensities corresponding to the inter-band scattering, which is 
similar to Fig.~\hyperref[fig3]{3(b)}. The key difference between the two patterns is that the intensity in
Fig.~\hyperref[fig4]{4(a)} undergoes an additional sign change as $q_x$ switches its sign,
and it increases in magnitude as $|q_x|$ grows. This behavior arises because 
$F^{ss'}_{A,0x}$ is an odd fuction of both $k_x$ and $k_x'$. The 
feature of the $(0,y)$ pattern in Fig.~\hyperref[fig4]{4(b)} is analogous to that of $(0,x)$, 
with the $x$ and $y$ axes swapped. Note that the FT-LDOS in the ($0,y$) channel is absent in the Zeeman-splitting scenario.

Similarly, there are only imaginary parts for $(x,z)$ and $(y,z)$ channels. The factor $F^{ss'}_{A,xz}$ reads 
\begin{equation}
	F^{ss'}_{A,xz}=\frac{1}{2}[s\frac{-\lambda{k'}_x}{\sqrt{\lambda^2{k'}^2+J^2{k'}_x^2{k'}_y^2}}+s'\frac{\lambda k_x}{\sqrt{\lambda^2k^2+J^2k_x^2k_y^2}}],
\end{equation}
which is similar to Eq.\eqref{eq30}. Such that the ($x,z$) pattern of SOC scenario in Fig.~\hyperref[fig4]{4(c)} has some similar characteristics to the ($0,x$) pattern of Zeeman splitting scenario in Fig.~\hyperref[fig4]{3(a)}. However, the $(x,z)$ pattern is antisymmetric about the $y$-axis, differing from that in Fig.~\hyperref[fig4]{3(a)}. Scattering with wave vector  $\mathbf{q}$ nearly parallel to the $x$-axis, where both $k_x$ and ${k'}_x$ are large, results in stronger intensities.
The $(y,z)$ pattern in Fig.~\hyperref[fig4]{4(d)} follow the same behavior as its $(x,z)$ counterpart with the $x$ and $y$ axes exchanged. Note that channel ($y,z$) is absent in the Zeeman splitting scenario.

\section{\label{sec5}Conclusion}

In summary, we have performed a comprehensive study of QPI patterns in metallic altermagnets, focusing on three key scenarios: the pristine $d$-wave altermagnet, the altermagnet under Zeeman splitting, and the altermagnet with Rashba SOC. By employing the T-matrix formalism and analyzing the FT-LDOS, we revealed that QPI patterns offer a direct means of probing both the geometric and spin configurations of altermagnetic Fermi surfaces.
For the pristine $d$-wave altermagnet, the QPI patterns clearly display the altermagnetic Fermi surface as two ellipses with orthogonal major axes and oppositely oriented spins. The introduction of Zeeman splitting or SOC alters these patterns, producing petal-shaped contours that signify a Lifshitz transition of the Fermi surface and delicate modifications in the intensity distribution. Moreover, the analysis of different probe and scattering spin channels allows us to extract detailed information of the spin textures from the QPI patterns, distinguishing between real patterns associated with Zeeman splitting and the appearance of imaginary components introduced by SOC. These results reveal key differences in the spin configurations for Zeeman and SOC scenarios, which manifest as unique suppression or enhancement effects in the QPI patterns.
Our theoretical and numerical results show that the modulations in QPI patterns under varying physical conditions effectively capture both the geometric and spin properties of altermagnetic Fermi surfaces. These findings provide a strong foundation for experimental studies using spin-resolved STS to explore the distinctive characteristics of altermagnetic materials. 

{Finally, some remarks regarding the application scenarios of our theory and its experimental relevance are given below. Here, we consider a 2D $d$-wave altermagnet described by a low-energy effective Hamiltonian. In real materials, the warping effects may exist and modify the QPI patterns. The general method we adopt here is well-suited for analyzing the resulting effects arising from changes in the shape and spin texture of the Fermi surface. Extending our theory to 3D materials is an intriguing direction for future exploration. The challenge may arise from the complexity due to the presence of surface states and surface spin canting. For thin films of 3D altermagnets, STS measurements can still provide valuable insights into bulk states, provided the spin texture and density of states at the surface faithfully reflect the bulk properties. 
Moreover, we focus on the regime of classical magnetic impurity, which corresponds to magnetic
impurities with large spins~\cite{Balatsky2006}.  The opposite regime, involving
quantum magnetic impurities, and their interaction with altermagnets~\cite{Diniz2024}, is an interesting
question for future research. It can be expected that the spin dynamics of these impurities can significantly influence QPI patterns and introduce a nontrivial energy dependence, providing further insights into the material properties~\cite{Derry2015,Mitchell2015,Mitchell2013}. 
To observe the predicted QPI patterns using current spin-resolved STS technologies, several experimental conditions are necessary~\cite{Oka2014}. High spatial resolution is required to resolve fine details. Low-temperature and ultra-high vacuum environments are essential to minimize thermal effects and surface contamination, ensuring clear measurements. The spin polarization of the STS tip must also be carefully controlled to accurately detect spin-dependent signals.}

{\emph{Note added.} After submitting our manuscript, we became aware of two related works~\cite{Sukhachov2024,Chen2024}}.

\begin{acknowledgments}
W. C. acknowledges financial support from
the National Natural Science Foundation of
China (No. 12222406 and No. 12074172), the Natural Science Foundation of Jiangsu Province (No. BK20233001),
the Fundamental Research Funds for the Central Universities (No. 2024300415), and
the National Key Projects for Research and Development of China (No. 2022YFA120470).
\end{acknowledgments}

\appendix*
\renewcommand{\appendixname}{APPENDIX}
\section{\label{appendixA}Analysis of QPI patterns}
\renewcommand\theequation{A\arabic{equation}}
\setcounter{figure}{0}
\renewcommand{\thefigure}{A\arabic{figure}}

There are various combinations of $A_{s(s')}(\mathbf{k}+\mathbf{q},\omega)$, $A_{s(s')}(\mathbf{k},\omega)$, $B_{s(s')}(\mathbf{k}+\mathbf{q},\omega)$ and $B_{s(s')}(\mathbf{k}+\mathbf{q},\omega)$ in the response function $\Lambda_{\alpha\beta}(\mathbf{q},\omega)$ defined in Eq.~\eqref{eq18}.
We note that all the terms of the response function have the same singularities (omitting the spin-coherent factor). We start at the $BA$ type,
\begin{eqnarray}
&&-\frac{1}{\pi}\int\mathrm{d}^2\mathbf{k}B_{s}(\mathbf{k}+\mathbf{q},\omega)A_{s'}(\mathbf{k},\omega)\nonumber\\
&&=\int\mathrm{d}^2\mathbf{k} \mathcal{P}\frac{1}{\omega-\omega^{s}_{\mathbf{k}+\mathbf{q}}}\delta(\omega-\omega^{s'}_{\mathbf{k}})
\nonumber\\
&&=\oint_{\omega^{s'}_{\mathbf{k}}=\omega} \mathcal{P}\frac{1}{\omega-\omega^{s}_{\mathbf{k}+\mathbf{q}}} \frac{1}{|\nabla_{\mathbf{k}}\omega_{\mathbf{k}}^{s'}|}\mathrm{d}l
\end{eqnarray}
The integral reduces to a line integral with respect to arc length along the isoenergy surface $\omega^{s'}_{\mathbf{k}}=\omega$. We may parameterize the $n$-th isoenergy surface as $\mathbf{k}_{\omega}^{n,s'}(t)=[f_{\omega}^{n,s'}(t),g_{\omega}^{n,s'}(t)]$ with $t\in[0,2\pi)$. For brevity we would omit the index $n$. Then the Cauchy principal value
\begin{equation}\label{eqA2}
\int_0^{2\pi} \mathcal{P}\frac{1}{\omega-\omega^{s}_{\mathbf{k}+\mathbf{q}}(t)} \frac{\sqrt{\dot{f_{\omega}^{s'}}^2(t)+\dot{g_{\omega}^{s'}}^2(t)}}{|\nabla_{\mathbf{k}}\omega_{\mathbf{k}}^{s'}(t)|}\mathrm{d}t
\end{equation}
diverges (for given $\mathbf{q}$) only if
\begin{equation}
\exists t_0\in[0,2\pi), [\omega-\omega^{s}_{\mathbf{k}+\mathbf{q}}(t_0)=0] \wedge [\frac{\mathrm{d}\omega^{s}_{\mathbf{k}+\mathbf{q}}}{\mathrm{d}t}(t_0)=0].
\end{equation}
The first condition promises that the scattered quasiparticle is on the isoenergy surface, and the second one points out that $\mathbf{q}$ should be specified. For the $AB-$type $A_{s'}(\mathbf{k}+\mathbf{q})B_{s}(\mathbf{k})=B_{s}(\mathbf{k})A_{s'}(\mathbf{k}+\mathbf{q})$, we just substitute $\mathbf{k}$ as $\mathbf{k}-\mathbf{q}$ and the similar condition is given.

The autocorrelation of $BB$ type is typically dominated  by the $AA-$type one which is written as
\begin{eqnarray}\label{eqA4}
&&\int\mathrm{d}^2\mathbf{k}A_{s}(\mathbf{k}+\mathbf{q},\omega)A_{s'}(\mathbf{k},\omega)\nonumber\\
&&=\int\mathrm{d}^2\mathbf{k} \delta(\omega-\omega^{s}_{\mathbf{k}+\mathbf{q}})\delta(\omega-\omega^{s'}_{\mathbf{k}})
\nonumber\\
&&=\oint_{\omega^{s'}_{\mathbf{k}}=\omega} \delta(\omega-\omega^{s}_{\mathbf{k}+\mathbf{q}}) \frac{1}{|\nabla_{\mathbf{k}}\omega_{\mathbf{k}}^{s'}|}\mathrm{d}l
\nonumber\\
&&=\int_0^{2\pi}  \delta(\omega-\omega^{s}_{\mathbf{k}+\mathbf{q}}(t))\frac{\sqrt{\dot{f_{\omega}^{s'}}^2(t)+\dot{g_{\omega}^{s'}}^2(t)}}{|\nabla_{\mathbf{k}}\omega_{\mathbf{k}}^{s'}(t)|}\mathrm{d}t
\end{eqnarray}
with the same parameterization as Eq.~\eqref{eqA2}. Because of the property of Dirac's delta function, we have
\begin{equation}
\delta(\omega-\omega^{s}_{\mathbf{k}+\mathbf{q}}(t))=\sum_{t_0} \delta(t-t_0)\frac{1}{|\frac{\mathrm{d}}{\mathrm{d}t}\omega^{s}_{\mathbf{k}+\mathbf{q}}(t_0)|},
\end{equation}
where $t_0$ is the solution of $\omega-\omega^{s}_{\mathbf{k}+\mathbf{q}}(t)=0$. Such that the integral in Eq.~\eqref{eqA4} diverges only if
\begin{equation}
\exists t_0\in[0,2\pi), [\omega-\omega^{s}_{\mathbf{k}+\mathbf{q}}(t_0)=0] \wedge [\frac{\mathrm{d}\omega^{s}_{\mathbf{k}+\mathbf{q}}}{\mathrm{d}t}(t_0)=0].
\end{equation}

\begin{figure}[htbp]\label{figA1}
	\centering
	\includegraphics[scale=0.33]{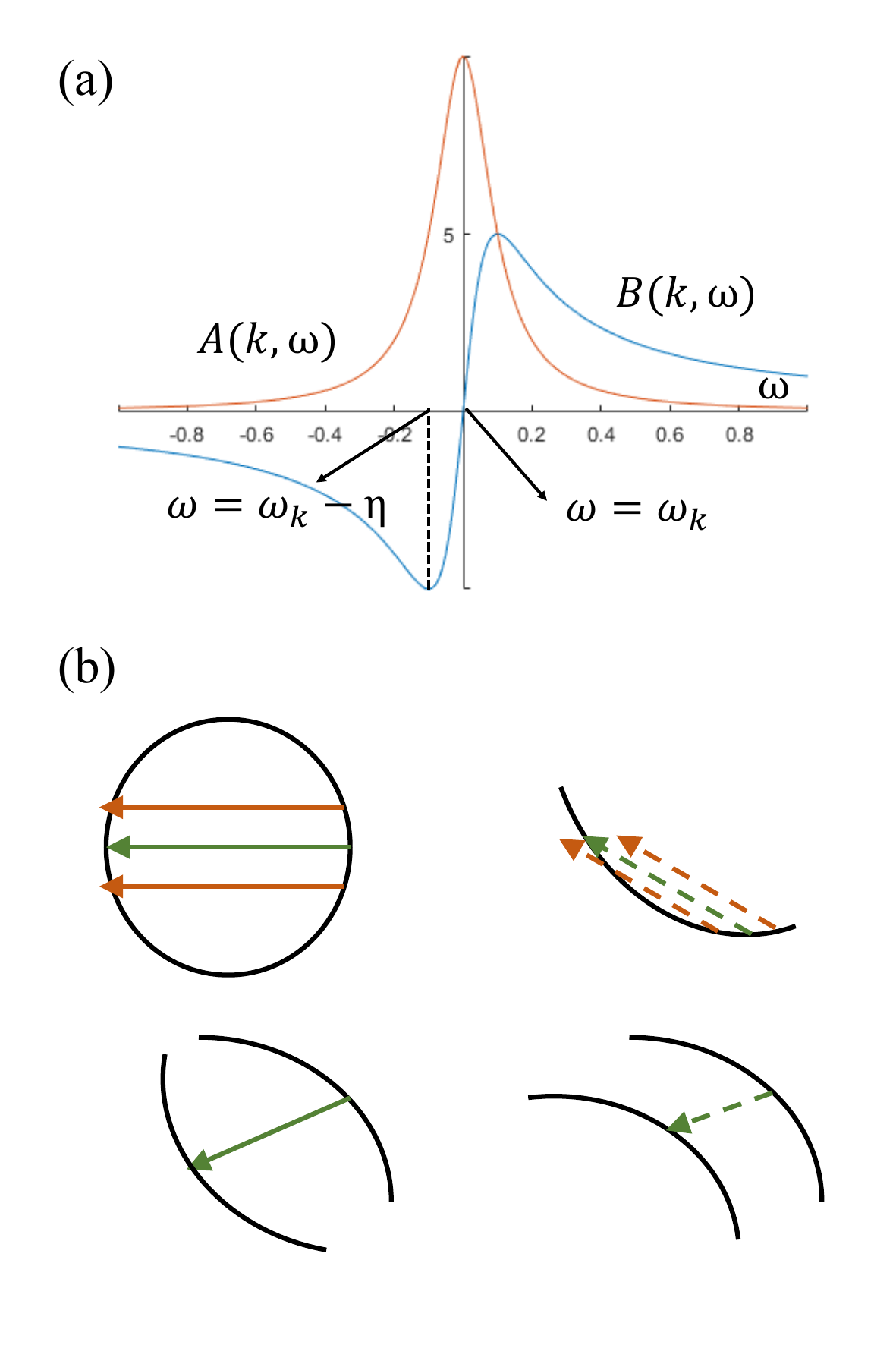}
	\caption{Stability of singularities for $AB$- and $BA$-type terms in the response function. (a) Amplitude of $A(\mathbf{k},\omega)$ and $B(\mathbf{k},\omega)$ as a function of $\omega$, considering a positive finite $\eta$ that models the finite lifetime of quasiparticles. (b) Examples of stable and unstable singularities for a flat isoenergy surface and near-nesting scattering. Stable singularities are indicated by solid lines, while unstable ones are shown with dashed lines. }
\end{figure}

We can see that all terms diverges under the same condition. Furthermore, it can be expressed intuitively by considering the $AA-$type (a.k.a. the joint density of states)
\begin{eqnarray}
&&\int\mathrm{d}^2\mathbf{k}A_{s}(\mathbf{k}+\mathbf{q},\omega)A_{s'}(\mathbf{k},\omega)\nonumber\\
&&=\int\mathrm{d}^2\mathbf{k} \delta(\omega-\omega^{s}_{\mathbf{k}+\mathbf{q}})\delta(\omega-\omega^{s'}_{\mathbf{k}})
\nonumber\\
&&=\int\mathrm{d}E\mathrm{d}E' \frac{\delta(\omega-E)\delta(\omega-E')}{|\frac{\partial(\omega^{s}_{\mathbf{k}+\mathbf{q}},\omega^{s'}_{\mathbf{k}})}{\partial(k_x,k_y)}|_{\mathbf{k}_{0}}}
\nonumber\\
&&=\sum_{\mathbf{k}_{0}}\frac{1}{|\nabla\omega_{\mathbf{k}_{0}+\mathbf{q}}^s\times \nabla\omega_{\mathbf{k}_{0}}^{s'}|},
\end{eqnarray}
where $\mathbf{k}_{0}$ is the solution of the equations
\begin{eqnarray}
\omega-\omega^{s}_{\mathbf{k}_{0}+\mathbf{q}}=0,\ \
\omega-\omega^{s'}_{\mathbf{k}_{0}}=0.
\end{eqnarray}
$\nabla\omega_{\mathbf{k}_{0}+\mathbf{q}}^s=[\partial_{k_x}\omega^{s}_{\mathbf{k}_{0}+\mathbf{q}},\partial_{k_y}\omega^{s}_{\mathbf{k}_{0}+\mathbf{q}}]$ and $\nabla\omega_{\mathbf{k}_{0}}^{s'}=[\partial_{k_x}\omega^{s'}_{\mathbf{k}_{0}},\partial_{k_y}\omega^{s'}_{\mathbf{k}_{0}}]$ are the normal vectors of isoenergy surface at $\mathbf{k}_{0}+\mathbf{q}$ and $\mathbf{k}_{0}$, respectively. Whereupon we note that the divergence point $\mathbf{q}$ of the response function $\Lambda_{\alpha\beta}(\mathbf{q},\omega)$ is determined by two points $\mathbf{k}_{0}+\mathbf{q}$ and $\mathbf{k}_{0}$ on the isoenergy surface whose normal vectors are parallel.

The spin-coherent factor defined in Eq.~\eqref{eq17} for a generic two-band Hamiltonian $H_0(\mathbf{k})=E_0(\mathbf{k})+\mathbf{d}(\mathbf{k})\cdot\boldsymbol{\sigma}$ reads,
\begin{eqnarray}\label{eqA9}
&&F_{\alpha\beta}^{ss'}(\mathbf{k},\mathbf{k}+\mathbf{q})\nonumber\\
&&=\frac{1}{4}\mathrm{Tr}[\sigma_\alpha\sigma_{\beta}+s\sum_{i}\sigma_\alpha\sigma_i\sigma_{\beta}\hat{\mathbf{d}}_{\mathbf{k}+\mathbf{q}}^i\nonumber\\
&&+s'\sum_{i}\sigma_\alpha\sigma_{\beta}\sigma_i\hat{\mathbf{d}}_{\mathbf{k}}^i+ss'\sum_{ij}\sigma_\alpha\sigma_i\sigma_{\beta}\sigma_j\hat{\mathbf{d}}_{\mathbf{k}+\mathbf{q}}^i\hat{\mathbf{d}}_{\mathbf{k}}^j],
\end{eqnarray}
where $j,l=x,y,z$, $\alpha,\beta=0,x,y,z$ and $\hat{\mathbf{d}}_{\mathbf{k}}=\frac{1}{\sqrt{\sum_i (\mathbf{d}_{\mathbf{k}}^i )^2}}[\mathbf{d}_{\mathbf{k}}^x,\mathbf{d}_{\mathbf{k}}^y,\mathbf{d}_{\mathbf{k}}^z]$. Here we exploit the identity $|s,\mathbf{k}\rangle\langle s,\mathbf{k}|=\frac{1}{2}(1+s\hat{\mathbf{d}}_{\mathbf{k}}\cdot\boldsymbol{\sigma})$. For different $\alpha$ and $\beta$, it can be divided into these four cases:
\begin{itemize}
\item [1)]
When $\alpha=\beta$,
\begin{equation}
F_{\alpha\alpha}^{ss'}(\mathbf{k},\mathbf{k}+\mathbf{q})=\frac{1}{2}(1+ss'\hat{\mathbf{d}}_{\mathbf{k}+\mathbf{q}}\cdot \hat{R}_\alpha\hat{\mathbf{d}}_{\mathbf{k}}).
\end{equation}
Here $\hat{R}_\alpha$ is a $\pi$-rotation about the $\alpha=x,y,z$ axis and $\hat{R}_0=\hat{\mathbb{I}}$.
\item [2)]
When $\alpha=0$,$\beta\neq0$,
\begin{eqnarray}\label{eqA11}
&&F_{0\beta}^{ss'}(\mathbf{k},\mathbf{k}+\mathbf{q})\nonumber\\
&&=\frac{1}{2}(s\hat{\mathbf{d}}_{\mathbf{k}+\mathbf{q}}^{\beta}+s'\hat{\mathbf{d}}_{\mathbf{k}}^{\beta}+iss'\sum_{ij}\epsilon_{\beta ji}\hat{\mathbf{d}}_{\mathbf{k}+\mathbf{q}}^i\hat{\mathbf{d}}_{\mathbf{k}}^j)\nonumber\\
&&=\frac{1}{2}[s\hat{\mathbf{d}}_{\mathbf{k}+\mathbf{q}}^{\beta}+s'\hat{\mathbf{d}}_{\mathbf{k}}^{\beta}+iss'(\hat{\mathbf{d}}_{\mathbf{k}}\times\hat{\mathbf{d}}_{\mathbf{k}+\mathbf{q}})^{\beta}].
\end{eqnarray}

\item [3)]
When $\beta=0$,$\alpha\neq0$,
\begin{eqnarray}
&&F_{\alpha0}^{ss'}(\mathbf{k},\mathbf{k}+\mathbf{q})\nonumber\\
&&=\frac{1}{2}(s\hat{\mathbf{d}}_{\mathbf{k}+\mathbf{q}}^{\alpha}+s'\hat{\mathbf{d}}_{\mathbf{k}}^{\alpha}+iss'\sum_{ij}\epsilon_{\alpha ij}\hat{\mathbf{d}}_{\mathbf{k}+\mathbf{q}}^i\hat{\mathbf{d}}_{\mathbf{k}}^j)\nonumber\\
&&=\frac{1}{2}[s\hat{\mathbf{d}}_{\mathbf{k}+\mathbf{q}}^{\alpha}+s'\hat{\mathbf{d}}_{\mathbf{k}}^{\alpha}+iss'(\hat{\mathbf{d}}_{\mathbf{k}+\mathbf{q}}\times\hat{\mathbf{d}}_{\mathbf{k}})^{\alpha}].
\end{eqnarray}

\item [4)]
When $\alpha\neq0$, $\beta\neq0$, $\alpha\neq\beta$,
\begin{eqnarray}
&&F_{\alpha\beta}^{ss'}(\mathbf{k},\mathbf{k}+\mathbf{q})\nonumber\\
&&=\frac{1}{2}[is\sum_i\epsilon_{\alpha i\beta}\hat{\mathbf{d}}_{\mathbf{k}+\mathbf{q}}^i+is'\sum_i\epsilon_{\alpha \beta i}\hat{\mathbf{d}}_{\mathbf{k}}^i\nonumber\\
&&+ss'(\hat{\mathbf{d}}_{\mathbf{k}+\mathbf{q}}^\alpha\hat{\mathbf{d}}_{\mathbf{k}}^{\beta}+\hat{\mathbf{d}}_{\mathbf{k}+\mathbf{q}}^{\beta}\hat{\mathbf{d}}_{\mathbf{k}}^{\alpha})].
\end{eqnarray}

\end{itemize}

Finally, we consider the stability of singularities in the response functions when accounting for the finite lifetime of quasiparticles \cite{Zhang2019}. In the Green's functions, we replace the infinitesimal term $0^+$ with a positive finite value $\eta$, which models the finite lifetime of quasiparticles. The key argument is that the singularities associated with $AB$ and $BA$ types, as defined in Eq.~\eqref{eq18}, become unstable when the isoenergy surface is approximately flat over a certain length. Notably, $B(\mathbf{k},\omega)$ and $A(\mathbf{k},\omega)$ behave differently around $\omega=\omega_{\mathbf{k}}$ (see Fig.\hyperref[figA1]{A1(a)}). Specifically, $B(\mathbf{k},\omega)$ vanishes at $\omega=\omega_{\mathbf{k}}$ and changes sign on either side. The response function exhibits significant weighting when $|\omega-\omega_{\mathbf{k}}|<\eta$. For large-momentum backscattering (illustrated in Fig.\hyperref[figA1]{A1(b)}), the scattered-out quasiparticle consistently has an energy $\omega > \omega_{\mathbf{k}}$, resulting in a positive contribution to the $AB$-type response function. However, when scattering occurs on an approximately flat contour (relative to $\mathbf{q}$) or near a nesting contour, both $\omega > \omega_{\mathbf{k}}$ and $\omega < \omega_{\mathbf{k}}$ quasiparticles contribute to the $AB$-type response function, leading to a cancellation of these contributions. In contrast, the joint density of states, which is the autocorrelation of $A(\mathbf{k},\omega)$, always receives positive contributions, primarily from around $\omega = \omega_{\mathbf{k}}$. Consequently, the QPI patterns resulting from $AB$ and $AA$ types differ for quasiparticles with finite lifetimes.

\end{document}